\documentclass[manuscript,screen]{acmart}
\AtBeginDocument{%
  \providecommand\BibTeX{{%
    Bib\TeX}}}

\setcopyright{acmlicensed}
\copyrightyear{2018}
\acmYear{2018}
\acmDOI{XXXXXXX.XXXXXXX}

\acmConference[Conference acronym 'XX]{Make sure to enter the correct
  conference title from your rights confirmation emai}{June 03--05,
  2018}{Woodstock, NY}
\acmISBN{978-1-4503-XXXX-X/18/06}

\usepackage{amsmath}
  
\usepackage{multicol}
\usepackage{multirow}
\usepackage{amssymb,amsfonts}
\usepackage{algorithmic}
\usepackage{subfigure}
\usepackage{graphicx}
\usepackage{textcomp}
\usepackage{makecell}
\usepackage{xcolor}
\usepackage{url}
\usepackage{colortbl}
\definecolor{green_best}{HTML}{009724}
\usepackage[braket, qm]{qcircuit} 
\def\BibTeX{{\rm B\kern-.05em{\sc i\kern-.025em b}\kern-.08em
    T\kern-.1667em\lower.7ex\hbox{E}\kern-.125emX}}
\definecolor{Celadon}{RGB}{175, 225, 175}
\begin{document}

\title{Optimal Toffoli-Depth Quantum Adder}

\author{Siyi Wang}
\email{siyi002@e.ntu.edu.sg}
\orcid{0009-0006-9128-1857}
\affiliation{%
  \institution{Nanyang Technological University}
  \country{Singapore}
  \postcode{43017-6221}
}

\author{Suman Deb}
\affiliation{%
 \institution{Nanyang Technological University}
 \country{Singapore}}
 
\author{Ankit Mondal}
\affiliation{%
  \institution{Indian Institute of Technology Delhi}
  \city{New Delhi}
  \country{India}
}

\author{Anupam Chattopadhyay}
\affiliation{%
 \institution{Nanyang Technological University}
 \country{Singapore}}

\renewcommand{\shortauthors}{Wang et al.}

\begin{abstract}
Efficient quantum arithmetic circuits are commonly found in numerous quantum algorithms of practical significance. Till date, the logarithmic-depth quantum adders includes a constant coefficient $k \geq 2$ while achieving the Toffoli-Depth of $k\log{}n + \mathcal{O}(1)$. In this work, 160 alternative compositions of the carry-propagation structure are comprehensively explored to determine the optimal depth structure for a quantum adder. By extensively studying these structures, it is shown that an exact Toffoli-Depth of $\log{}n + \mathcal{O}(1)$ is achievable. This presents a reduction of Toffoli-Depth by almost $50\%$ compared to the best known quantum adder circuits presented till date. We demonstrate a further possible design by incorporating a different expansion of propagate and generate forms, as well as an extension of the modular framework. Our paper elaborates on these designs, supported by detailed theoretical analyses and simulation-based studies, firmly substantiating our claims of optimality. The results also mirror similar improvements, recently reported in classical adder circuit complexity.

\end{abstract}

\ccsdesc[500]{Computer systems organization~Quantum computing}

\keywords{Quantum Computing,
Quantum Arithmetic,
T-depth Optimization,
Carry-lookahead Adder,
Sklansky Tree}


\maketitle
\section{Introduction}
Quantum computers powered by quantum algorithms promise to offer problem-solving abilities that will far exceed the most powerful supercomputers of today. These quantum algorithms rely on efficient arithmetic circuits to harness their full potential. While the quantum computing space continues to bustle with diverse research activities, optimizing quantum circuits remains a fundamental necessity. Quantum adders stand out as one of the key circuits in any quantum computing system. For instance, Shor's algorithm~\cite{Shor}, which can factorize large integers exponentially faster than a classical computer, relies on efficient quantum addition.

While adder circuits have been frequently explored in classical computing, they remain relatively less-chartered in the quantum computing paradigm. Arithmetic circuits have been extensively explored in classical computing. Interestingly, the Sklansky Tree~\cite{Sklansky} is widely applied for its minimal depth among all the other parallel prefix structures. However, implementing this structure in the quantum world presents significant challenges, primarily due to qubit non-copyability.

This paper proposes a novel quantum adder based on the Sklansky Tree, which is verified as the optimal depth structure among all the quantum adders.
Throughout this research, we faced a series of challenges. Specifically, the inability to copy qubits presented a barrier to the direct application of the prefix tree adders in the quantum world. To address this issue, we developed a quantum repeat gate that helps to efficiently incorporate the different Prefix Tree structures such as Sklansky~\cite{Sklansky} into the quantum world under specific constraints. The proposed repeat gate not only contributes to the establishment of an optimal-depth quantum adder but also effectively addresses the problems arising from qubit non-copyability in other specific quantum circuits.

Our primary contribution is the development of the optimal-depth quantum adder that achieves an remarkable Toffoli-Depth of $\log n + \mathcal{O}(1)$ for $n$ bit-sized additions. Obviously, it marks an important improvement over all the previous quantum carry-lookahead adders based on the Brent-Kung tree, which required a minimum of $2\log n + \mathcal{O}(1)$ Toffoli-Depth. By conducting a thorough assessment involving Toffoli-Depth, Qubit Count, and Toffoli-Count, this paper offers significant insights into the strengths and constraints of the quantum optimal-depth adder, enhancing the continuous development of quantum computing, which parallels the development of classical adder analysis \cite{Best_Classical_Adder}.
In brief, the main contributions include:
\begin{itemize}
\item Thoroughly investigate the prefix tree and explore computation forms using different propagation and generation techniques.
\item Propose innovative designs for optimal Toffoli-Depth adders, optimal Toffoli-Depth Ling adder and optimal Toffoli-Depth modular adder, achieving the peak performance in quantum addition circuits.
\end{itemize}

The rest of this paper is organized as follows: Section \ref{sec: Prev} describes the relevant previous research. In Section \ref{sec: Stru}, we introduce and compare the Sklansky Tree and other parallel prefix structures. Section \ref{sec: Desi} presents the overall design of the proposed optimal Toffoli-Depth quantum adder, Ling expansion and modular adder expansion.
Additionally, Section \ref{sec: Resl} offers a detailed performance analysis, including a comparison with dominant quantum adder designs. We conclude the paper and discuss future research directions in Section \ref{sec: Conc}.

\section{Related Work}\label{sec: Prev}
%
We provide a comprehensive overview of the related literature in this section. 
In the quantum world, reducing the Toffoli gate is a prominent trend~\cite{Payoff_Function}\cite{divider} due to its time-intensive property. Among all the different types of quantum adders, Carry-Lookahead Adders (CLAs) are specially designed to achieve reduced depth. In 2004, Draper et al. \cite{Draper} proposed a logarithmic-depth quantum carry-lookahead adder, reducing Toffoli-Depth from $O(n)$ to $O(\log n)$, resulting in notable efficiency improvements.
Subsequently, there was an obvious increase in the development of quantum CLAs. For example, Takahashi \textit{et al.} further optimized Draper's work and designed some important quantum CLAs~\cite{Takahashi08, Takahashi09}.
Furthermore, Wang et al.~\cite{wang2023higher} compared all the quantum CLAs and found Brent-Kung \cite{BK} to be the top choice for implementing quantum adders.

However, it is interesting to note that, while majority of the quantum CLAs are based on the Brent-Kung structure, the Kogge-Stone~\cite{KSTree} and Slansky~\cite{Sklansky} architectures  continue to have the lowest depth among all the parallel prefix structures in theory. 
For example, compared to Brent-Kung's $2\log n$ depth,  the Kogge-Stone and Slansky structures achieve n-bit addition with only half the depth.
However, research on the quantum prefix tree structure is currently limited. This is primarily due to the nature of the prefix tree structures which involve extensive bit-sharing operations. In classical computing, multiple operations that share the same input bit are possible simultaneously. In the quantum domain, due to the non-copyability of qubits, operations sharing the same input must be executed in separate time slots. As a result, previous works have favored the use of the Brent-Kung tree in quantum computing, considering it a more convenient and cost-effective approach.
Nonetheless, we have proposed a novel addition framework to solve this qubit-sharing challenge within specific constraints. Following a thorough comparison, we have found that the Sklansky Tree emerges as the optimal structure among all the quantum prefix tree adders.

In this paper, we propose an innovative quantum optimal-depth adder based on the Sklansky tree. Through systematic comparison with other existing designs, our adder demonstrates a Toffoli-Depth of $\log(n)+ \mathcal{O}(1)$, which significantly outperforms the Toffoli-Depth of previous Brent-Kung-based adders. Consequently, our design has the potential to become a noteworthy candidate for constructing large-scale quantum circuits, mirroring an advancement similar to classical adder circuit design~\cite{Best_Classical_Adder}.

\section{Design Choices for Depth Reduction}\label{sec: Stru}
\begin{figure}[ht!]
\centering
\includegraphics[width=0.99\linewidth]{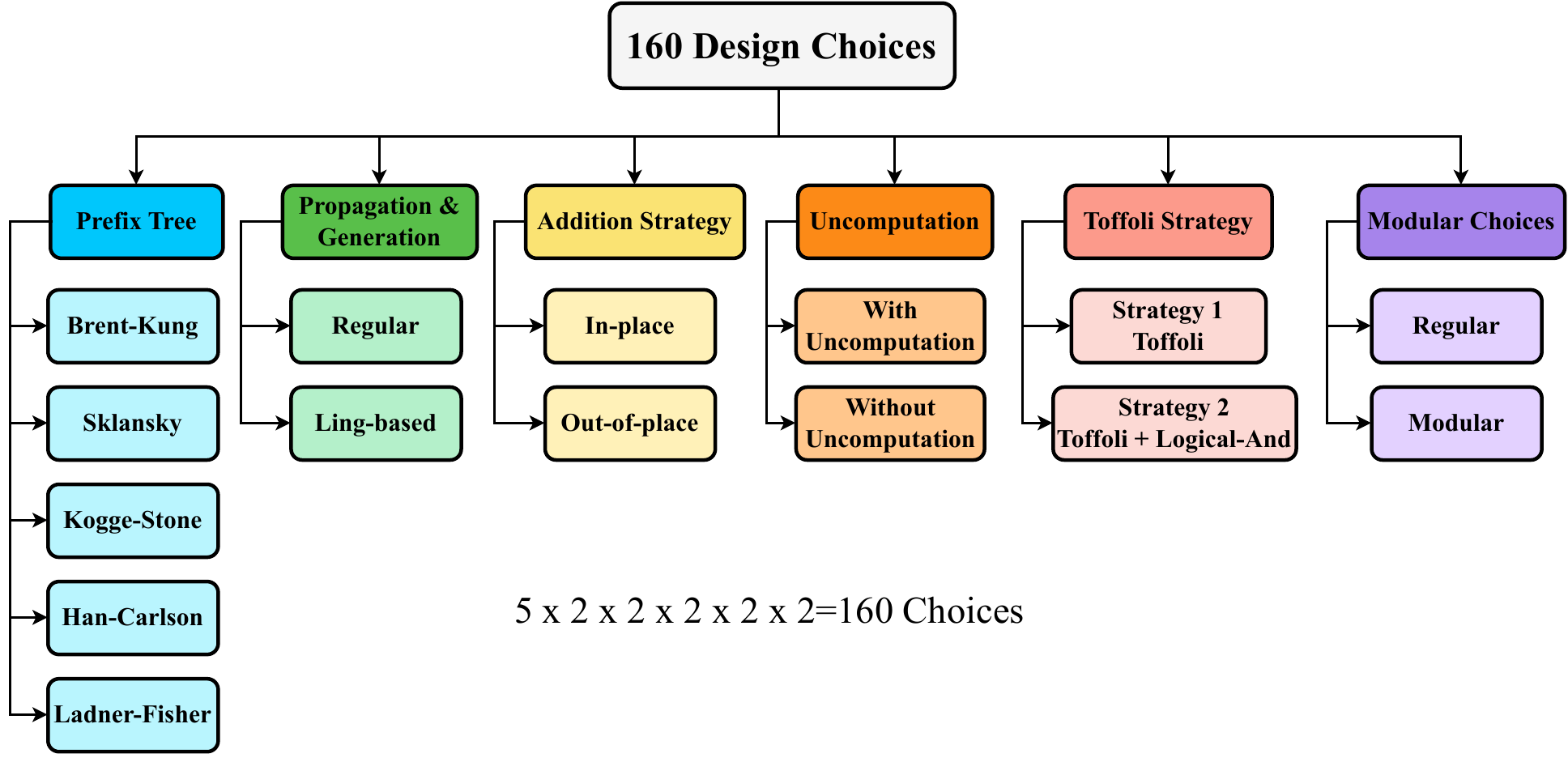}  
\caption{Design Choices.\label{fig:160Choices} }
\end{figure}

In this section, we elaborate on the choices to be made with the primary objective of reducing Toffoli-Depth.
As shown in Figure \ref{fig:160Choices}, in this study, we explored a total of 160 design options. In Sections \ref{sec:tree} and \ref{sec:PG}, we provided detailed descriptions of the Prefix tree and the Propagation and Generation Computation methods, respectively. Regarding addition strategies, there are two options: in-place addition and out-of-place addition, but this work primarily focuses on the quantum out-of-place adder with lower Toffoli-Depth. In Section \ref{sec:Uncomputation}, we discuss the uncomputation strategy, presenting a framework for an adder with uncomputation. Specifically, this framework is outlined by removing all uncomputation steps, such as step 3, to obtain a version without uncomputation. In Section \ref{sec:Toffoli}, two different Toffoli strategies are proposed. Specifically, Strategy 1 involves only Toffoli gates, while Strategy 2 combines Toffoli gates with logical-AND for further optimization of our circuits. For modular choices, we include both regular adders and modular adders, with specific details discussed in Section \ref{sec:Modular}.

\subsection{Choice of Prefix Tree\label{sec:tree}}
As shown in Table \ref{tab:table_Compare_prefix}, the classical Sklansky tree and Kogge-Stone tree stand out as the most efficient parallel prefix structures in terms of logical depth compared to other similar structures, such as the Brent-Kung tree and the Han-Carlson tree.

\begin{table}[ht!]
\centering
\caption{Comparison of dominant prefix trees for $n$-bit addition.
\label{tab:table_Compare_prefix}}
\begin{tabular}{|c|c|c|}
\hline \textbf{Prefix Tree} & \textbf{Logical depth} & \textbf{Operation nodes} \\\hline
 
 Brent-Kung &$2\log n -2$ &$2n-\log n -2$ \\\hline

 Sklansky &\textcolor{green_best}{$\log n$} &$(n/2)*\log n$ \\\hline
 
 Kogge-Stone &\textcolor{green_best}{$\log n$} &$n\log {(n/2)}+1$\\\hline 

  Han-Carlson &$\log n +1$ &$(n/2)*\log n$ \\\hline

 Ladner-Fisher &$\log n+1$ &$\frac{3}{4}n-1+(n/4)*\log n$  \\\hline

 
\end{tabular}
\end{table}

\begin{figure}[ht!]
    \centering
    \subfigure[Quantum Kogge-Stone.\label{fig:Q_KS}]{\includegraphics[width=0.48\linewidth]{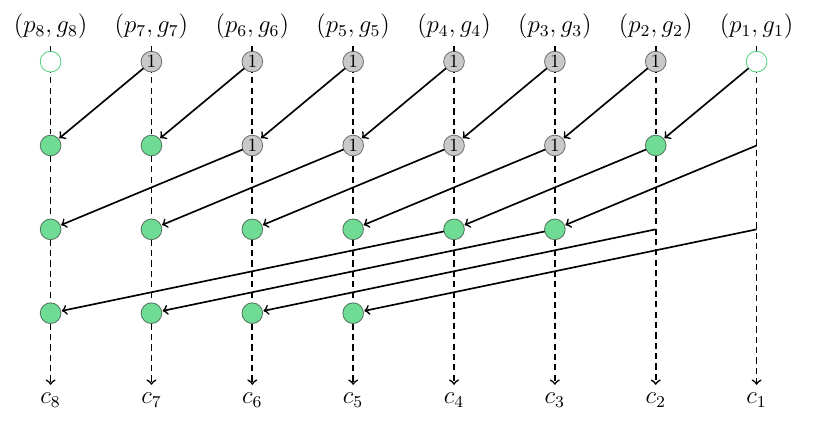}
    }
    \hfill
    \subfigure[Quantum Slansky. \label{fig:Q_Slan}]{\includegraphics[width=0.48\linewidth]{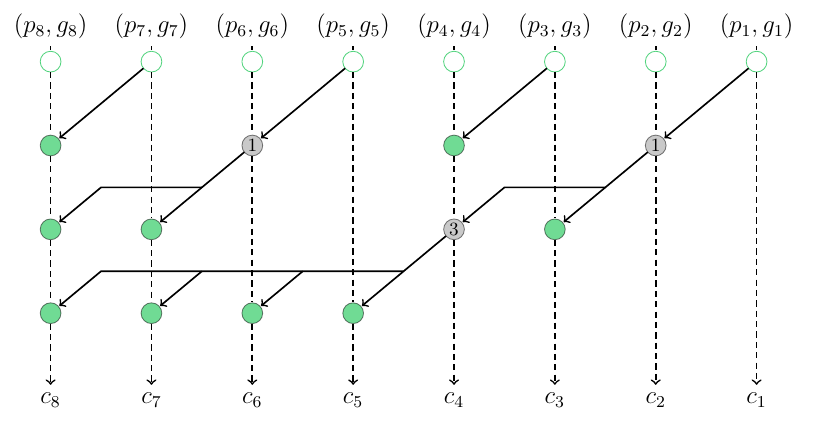}
    }
    \hfill
    \subfigure[Quantum Han-Carlson.\label{fig:Q_HC}]{\includegraphics[width=0.48\linewidth]{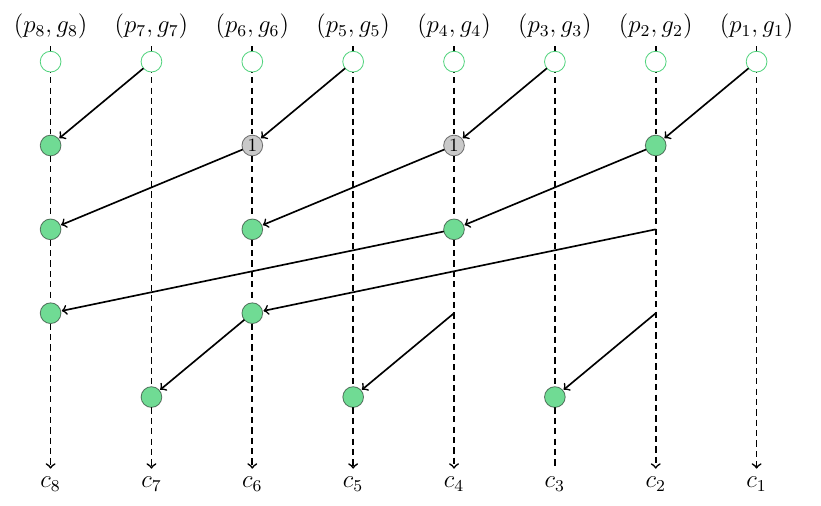}
    }
    \hfill
    \subfigure[Quantum Ladner-Fischer.\label{fig:Q_LF}]{
    \includegraphics[width=.48\linewidth]{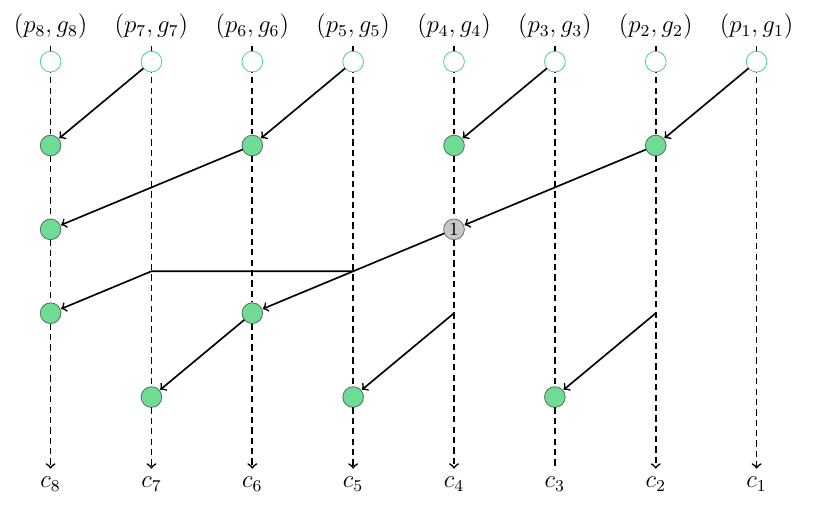}
    }
    \subfigure[Quantum Brent-Kung. \label{fig:Q_BK}]{\includegraphics[width=0.48\linewidth]{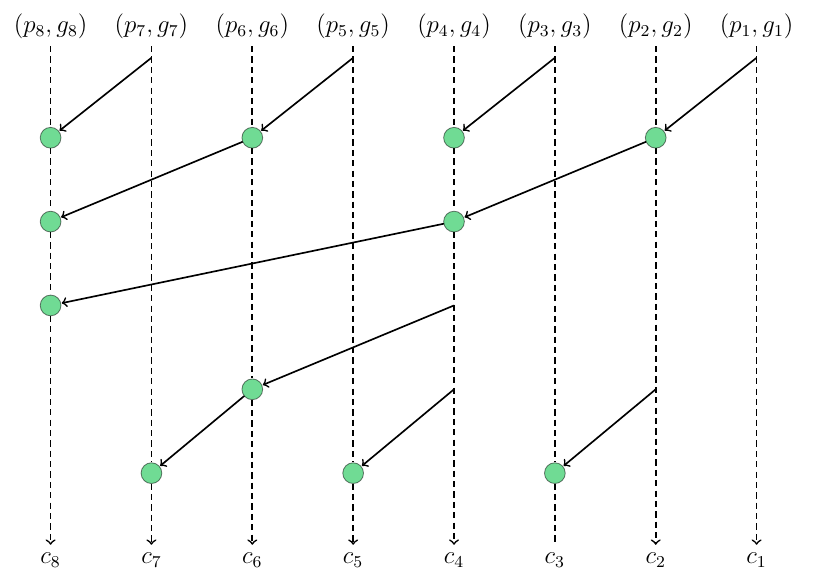}
    }
    \hfill
    \caption{Quantum Prefix Tree. Gray nodes in the prefix trees represent CNOT operations, with numbers in circles indicating the times of CNOT operations.\label{fig: Quantum_Prefix_Tree}}
\end{figure}

Among state-of-the-art logarithmic-depth quantum adders, the common approach is to use the Brent-Kung tree \cite{Draper,wang2023higher,wang2023reducing}. However, it is well known in the classical adder that the Kogge-Stone and Slansky prefix trees can achieve a significantly lower logical depth, i.e., $\log n$, reducing depth of the Brent-Kung tree by half. Given this substantial difference in logical depth, it is surprising that these prefix trees were not studied earlier. One possible reason for the same is the prefix tree incurs an overhead due to copy operation, which is inherently prevented in quantum domain due to no-cloning theorem. However, this limitation can be bypassed through introduction of ancilla qubits with initialization states as $\ket{1}$ or $\ket{0}$ and proper application of CNOT gate as shown in Figure \ref{fig: Quantum_Prefix_Tree}. 

Among the five prominent quantum prefix tree structures, Quantum Slansky (Figure \ref{fig:Q_Slan}) stands out as particularly advantageous. While it exhibits the lowest logical depth, Quantum Slansky also features significantly fewer CNOT operations compared to Quantum Koggle Stone, which has an equivalent logical depth.

\subsection{Choice of Propagate and Generate Computation\label{sec:PG}}

The efficiency of quantum carry-lookahead adder depends on the choice of propagate and Generate strategy. Here, two common approaches are described as follows:
\begin{itemize}
    \item \textbf{Conventional Propagation and Generation.} The conventional method employs standard logical operations for carry propagation and generation as shown in formulas \ref{e:Step2-1},\ref{e:Step2-2},\ref{e:Step2-3},\ref{e:calculate-c-1} and \ref{e:calculate-c-2}. 
\begin{eqnarray}
&(G_{0:0}, P_{0:0})&=(g_0, p_0)\label{e:Step2-1}\\
&(G_{0:i}, P_{0:i})&=(g_i, p_i)\circ (G_{0:i-1}, P_{0:i-1})\label{e:Step2-2}\\
\nonumber
&(g_x, p_x) \circ (g_y, p_y)&=(g_x+p_x\cdot g_y, p_x \cdot p_y)\\
&\quad&=(g_x\oplus (p_x\cdot g_y), p_x \cdot p_y)\label{e:Step2-3}\\
&c_0&=0\label{e:calculate-c-1}\\
&c_{i+1}&=g_i+p_i\cdot c_i=G_{0:i}\label{e:calculate-c-2}
\end{eqnarray}

It is widely used for its simplicity, providing a clear and straightforward construction of quantum carry-lookahead adders.
    \item \textbf{Ling-Based Propagation and Generation.} This method is represented by the Quantum Ling adder \cite{wang2023reducing},  which incorporates the Ling basis\cite{Ling} into the dominant Quantum Carry Look-Ahead Adders(CLA) which mainly based on Brent-Kung structure. The Ling-based propagation and generation structure introduces a more complex pre- calculation part, thereby incorporating an additional parallel calculation tree in the calculation process, resulting in a reduction in depth cost.
    Compared to the traditional approach, the complexity of Toffoli-Depth can be further reduced to $O(\log \frac{n}{2})$ from $O(\log n)$ by integrating the Ling expansion and utilizing a slightly different pre-calculation structure.
\end{itemize}

\section{The Design of Quantum Prefix Tree Adder}\label{sec: Desi}

In this section, a detailed description of our quantum prefix tree adders is provided.

\subsection{Evaluation Metrics}
The metrics used in this work are outlined as follows:
\begin{itemize}
    \item \textbf{Toffoli-Depth.} It represents the number of computational layers that include Toffoli gates, related to the overall time complexity. Toffoli-Depth can be converted to T-depth using specific decomposition methods, thereby allowing for the standardized evaluation.
    \item \textbf{Toffoli-Count.} It measures the total count of Toffoli gates within the circuit, providing an estimate of the gate complexity and quantum resource consumption. By using specific decomposition methods, it is convertible to T-count for standardized comparisons.
    \item \textbf{Qubit-Count.} It denotes the total number of qubits required, which correlates with the size of the quantum circuit.
\end{itemize}

These metrics enable a comprehensive evaluation of quantum circuit efficiency and practicality.

\subsection{Quantum Prefix Tree Adder\label{sec:Uncomputation}}

\begin{figure}[ht!]
    \centering
    \subfigure[Brent-Kung Quantum Adder.\label{fig:Step1}]{\includegraphics[width=0.49\linewidth]{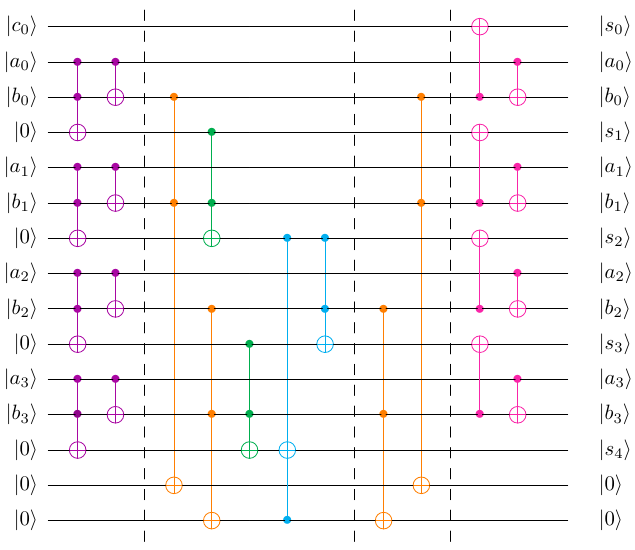}
    }
    \hfill
    \subfigure[Slansky Quantum Adder.\label{fig:Step1}]{\includegraphics[width=0.49\linewidth]{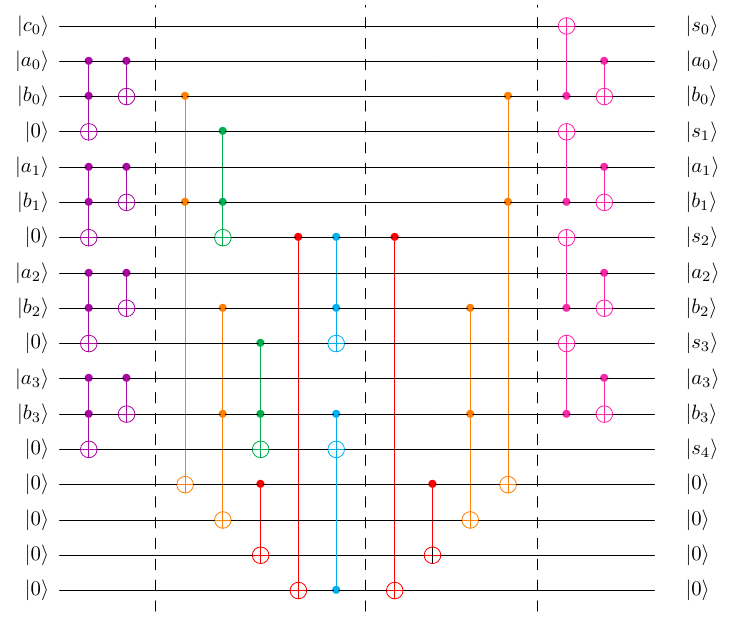}
    }
    \hfill
    \subfigure[Kogge-Stone Quantum Adder.\label{fig:Step1}]{\includegraphics[width=0.8\linewidth]{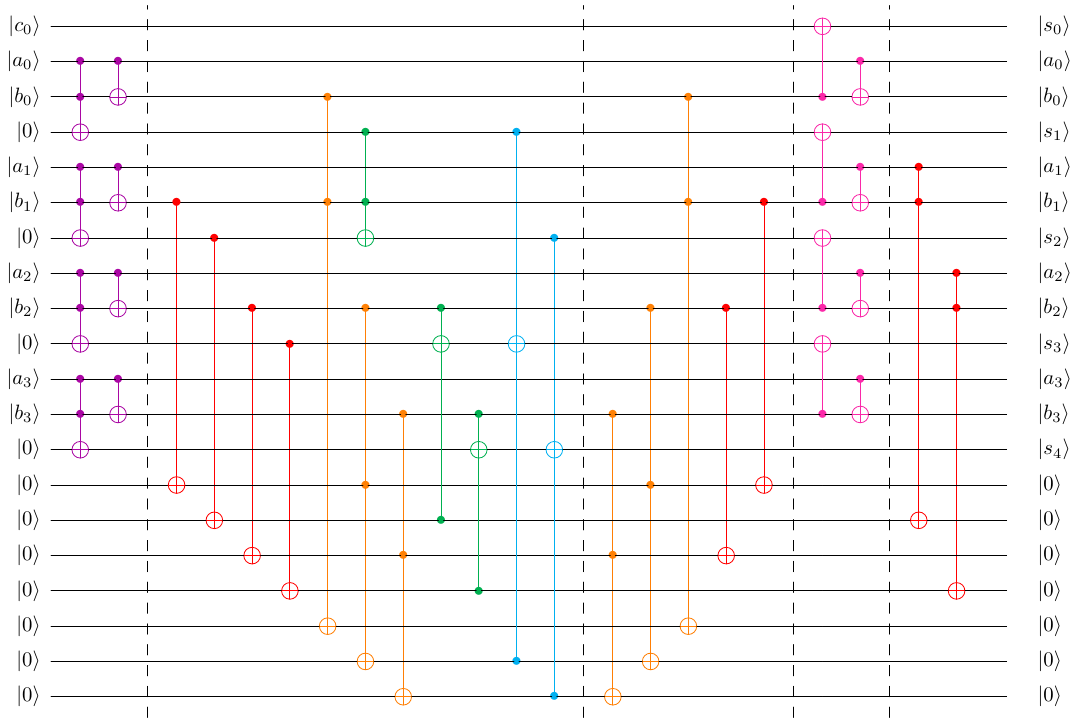}
    }
    \caption{4 bit Quantum Adder Examples\label{Overall_Design}. The dashed lines divide the entire circuit into five parts according to steps 1 to 4. In the first step, purple represents the calculation for initial propagation and generation. In the second step, red represents the copy operation, orange represents the first layer of propagation, and green and blue represent the first and the rest layers of generation, respectively. In the third step, orange represents propagation uncomputation, and red represents the uncomputation of the initial propagation copy $p_i$. Next, pink represents the operations of the fourth step. In the final step of Kogge-Stone Adder, we perform uncomputation on all the initially copied $g_i$, represented in red.}
\end{figure}

As shown in Figure \ref{Overall_Design}, we construct the Quantum prefix tree adder by combining the steps 1 to 4.
\begin{itemize}
    \item \textbf{Step 1.} The initial propagation $p_i$ and generation $g_i$ are calculated using a single ancilla qubit and one Toffoli gate. 
\begin{figure}[ht!]
\centering
\includegraphics[width=0.26\linewidth]{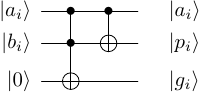} 
    \caption{Quantum circuit for $p_i$ and $g_i$ calculation.\label{fig:Step1}}
 \end{figure}
 
 Consequently, Toffoli-Depth is $1$, Toffoli-Count is $n$, and Qubit-Count is $(3n+1)$.
     \begin{eqnarray}
    \centering
    &p_i&= a_i \oplus b_i\label{e:Step1-1}\\
    &g_i&=a_i \cdot b_i\label{e:Step1-2}
    \end{eqnarray}

    \item \textbf{Step 2.} We implement both $P$ propagation and $G$ propagation (Figure \ref{fig: Step 2}) based on the corresponding quantum prefix tree structures as shown in Figure \ref{fig: 4-bit-Q-Prefix-Tree}.
    
    \begin{figure}[ht!]
    \centering
    \subfigure[Brent-Kung Prefix Tree.\label{fig:Step1}]{\includegraphics[width=0.3\linewidth]{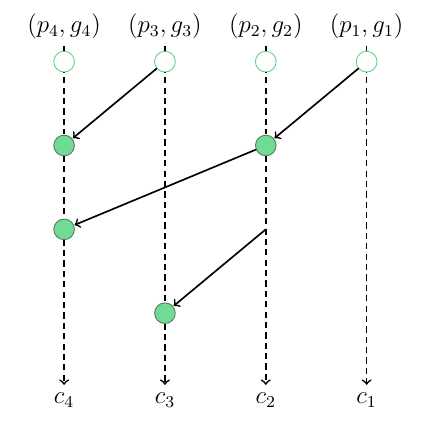}
    }
    \hfill
    \subfigure[Slansky Prefix Tree.\label{fig:Step1}]{\includegraphics[width=0.3\linewidth]{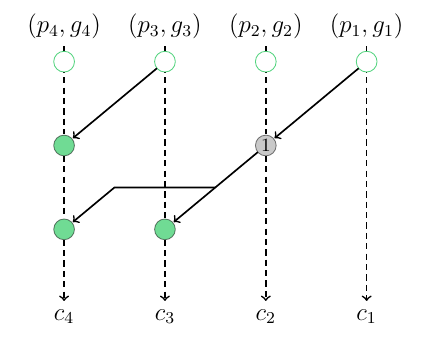}
    }
    \hfill
    \subfigure[Kogge-Stone Prefix Tree.\label{fig:Step1}]{\includegraphics[width=0.3\linewidth]{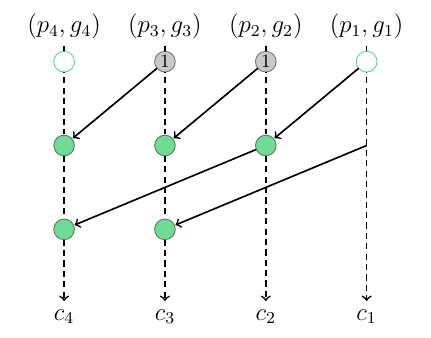}
    }
    \caption{4-bit Quantum Prefix Tree\label{fig: 4-bit-Q-Prefix-Tree}. Interestingly, the 4-bit Brent-Kung Prefix Tree is equivalent to the Han-Carlson and Ladner-Fischer structures.}
\end{figure}
\begin{figure}[ht!]
    \centering
    \includegraphics[width=.26\linewidth]{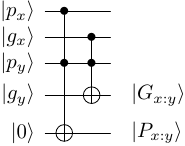}
    \caption{Step 2.  \label{fig: Step 2}}
\end{figure}
The cost of this step depends on the prefix tree structure employed. Specifically, for the quantum Kogge-Stone adder, this step involves a Toffoli-Depth of $(1 + \log n) $, a Toffoli-Count of $ (n\log n - \frac{3}{2}n + n\log{\frac{n}{2}} + 2) $, and an extra Qubit-Count of $ (3n\log n - \frac{7}{2}n + 5 )$.

    \item \textbf{Step 3.} In this step, $P$ and the copied $G$ are uncomputed based on different prefix tree structures. For example, considering the Kogge-Stone tree, this process incurs a Toffoli-Depth of $(\log n - 1 )$. The corresponding Toffoli-Count is $(2n\log n - \frac{7}{2}n + 5 )$, and no extra qubit.

    \item \textbf{Step 4.} This step involves sum calculation and $ p_i$ uncomputation without the employment of Toffoli gates. 
\begin{figure}[h!]
\centering
\includegraphics[width=0.29\linewidth]{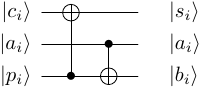}  
\caption{Quantum circuit for computing s from p and c\label{fig:Step3}}
\end{figure}

Besides, here is no extra qubit required.
\begin{eqnarray}
\centering
&s_i &= p_i \oplus c_i\label{e:Step3-1}\\
&s_{n+1} &=c_{n+1}\label{e:Step3-2}
\end{eqnarray}

\item \textbf{Additional Step.}  This step is exclusively designed for the quantum Kogge-Stone structure due to its intricate CNOT operations, necessitating an extra step to ensure the complete uncomputation of all ancillary qubits. 
As illustrated in Figure \ref{fig:Step5}, this step involves uncomputing all the copied $g_i$ using a Toffoli-Depth of 1 and a Toffoli-Count of $(n - 2)$. Similar to Step 4, this step does not require any ancilla.

\begin{figure}[ht!]
\centering
\includegraphics[width=0.26\linewidth]{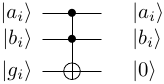}  
\caption{Quantum circuit for Uncomputation\label{fig:Step5} (only for copied $g_i$) }
\end{figure}    
\end{itemize}

The overall structures are depicted in Figure \ref{Overall_Design}, showcasing three distinct examples of 4-bit addition quantum circuits. Interestingly, the equivalence of quantum Brent-Kung, Han-Carlson, and Ladner-Fischer structures is evident within this proposed framework. The comprehensive cost analysis is provided in the subsequent section.

\section{Experimental Results and Discussion}\label{sec: Resl}

In the previous section, we proposed the general framework and specific quantum implementations of several quantum prefix tree adders. In this section, we will provide all the details of the results and discussions. Our investigation focuses on the quantum out-of-place adder due to its direct reflection of the prefix tree's influence and its lower Toffoli-Depth compared to the quantum In-place adder. Conversely, while the In-place adder follows a similar workflow, it typically involves more intricate supplementary steps, resulting in a deeper Toffoli-Depth.

\subsection{Identifying the Optimal Depth Adder Among Quantum Prefix Tree Adders\label{sec:Toffoli}}

First of all, we conducted a comprehensive analysis of the costs associated with all proposed quantum prefix tree adders to determine the optimal depth structure within the quantum realm. In this process, we introduced two distinct strategies.

\subsubsection{Strategy 1}
Under Strategy 1, we exclusively employ the Clifford+Toffoli gate set to describe and evaluate the entire quantum addition circuit. Remarkably, the Quantum Brent-Kung adder here is equivalent to the existing Draper out-of-place adder\cite{Draper}. 

\begin{table}[ht!]
\caption{Performance analysis of different quantum adders utilizing Strategy 1.\label{Table_Strategy1}}
\begin{tabular}{|c|c|c|c|}
\hline
\textbf{Adder} & \textbf{Toffoli-Count} & \textbf{Toffoli-Depth} & \textbf{Qubit-Count} \\ \hline
\textbf{Brent-Kung}&$5n-3\omega(n)-3\left \lfloor\log n\right \rfloor-1$ &$4+\left \lfloor\log n\right \rfloor+\left \lfloor\log \frac{n}3\right \rfloor$&$4n+1-\omega(n)-\left \lfloor\log n\right \rfloor$ \\ \hline
\textbf{Sklansky }              &$\frac{3}{2}n\log n +2\left \lceil \log n \right \rceil-n$&\textcolor{green_best}{$2\log n+1$}&$n+n\log n+\left \lceil \log n \right \rceil+2$\\ \hline
\textbf{Kogge-Stone}&$3n\log n+n\log {\frac{n}{2}}-3n+5$&$2\log n+2$&$3n\log n -\frac{n}{2}+6$\\ \hline
\textbf{Han-Carlson }           &$n+\frac{3}{2}n\log n -2\left \lfloor \frac{n}{2} \right \rfloor$&$2\log n+3$&$\frac{3}{2}n+n\log n-\left \lfloor \frac{n}{2} \right \rfloor+3$   \\ \hline
\textbf{Ladner-Fisher }         & $\frac{13n}{4}+\frac{3n\log n}{4}-2\left \lfloor \frac{n}{2} \right \rfloor-3$&$2\log n+3$& $3n+\frac{n\log n}{2}-\left \lfloor \frac{n}{2} \right \rfloor+1$\\ \hline
\end{tabular}
\end{table}

By applying strategy 1, the corresponding costs are detailed in Table \ref{Table_Strategy1}. Unfortunately, it is obvious that compared with the quantum Brent-Kung Prefix Tree adder, the alternative prefix tree structures we proposed do not demonstrate significant advantages, especially concerning the Toffoli-Depth.

\subsubsection{Strategy 2}
In 2018, Gidney proposed a more cost-effective logical AND structure for implementing a pair of Toffoli gates \cite{Gidney2018halvingcostof}. By utilizing this design, Strategy 2 aims to further optimize quantum adders based on Strategy 1. Specifically, for the proposed quantum prefix tree adder framework, only the P propagation involves Toffoli pairs. Hence, we utilize the logical AND structure for all Toffoli gates in both the P computation and P uncomputation stages. For other unpaired Toffoli gates, we still use Clifford+Toffoli gate set to describe them.

\begin{figure}[!ht]{
    \centering
    \subfigure[Computation\label{fig:Logical-And: Computation}]{ 
\includegraphics[width=.62\linewidth]{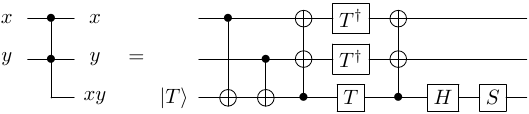}  
}
    \hspace{1.6cm}
\smallskip
    \centering
    \subfigure[Uncomputation\label{fig:Logical-And: Uncomputation}]{
    \includegraphics[width=.47\linewidth]{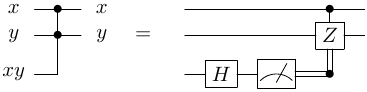
    } 
    }
    \caption{Gidney's Logical-And structure.\label{fig:Logical_And}}
}
\end{figure}

Under Strategy 2, the respective costs can be found in Table \ref{Compare_Strategy_2}. It is evident that all proposed designs significantly outperform the original Quantum Brent-Kung adder in terms of Toffoli-Depth. Particularly, the Quantum Slansky adder exhibits the quantum-optimal Toffoli-Depth, which is only $log(n)+1$. 

\begin{table}[ht!]
\centering
\caption{Performance analysis of different quantum adders utilizing Strategy 2.\label{Compare_Strategy_2}}
\begin{tabular}{|c|c|c|c|c|c|}
\hline
\textbf{Adder} & \textbf{Extra T Count} & \textbf{Toffoli Count} & \textbf{Extra T Depth} & \textbf{Toffoli Depth} & \textbf{Qubit Count} \\ \hline
\textbf{Brent-Kung }            & $8n-4\log n -8-4\left \lfloor\frac{n}{2}\right \rfloor$  & $2n-\log n -2 $                     &                        2&$2\log n-1$&$4n+1-\omega(n)-\left \lfloor\log n\right \rfloor$ \\ \hline
\textbf{Sklansky}               &$2n\log n-4n+4\left \lceil \log n \right \rceil)$&            $\frac{n\log n}{2}$ &                        2&\textcolor{green_best}{$\log n+1$} &$n+n\log n+\left \lceil \log n \right \rceil+2$\\ \hline
\textbf{Kogge-Stone }           &$8n\log n-14n+20$& $n\log {n}-1$ &                        2&$\log n+2$&$3n\log n -\frac{n}{2}+6$\\ \hline
\textbf{Han-Carlson }           &$2n\log n-4\left \lfloor \frac{n}{2} \right \rfloor$ &        $\frac{n\log n}{2}$&                        2&$\log n+2$&$\frac{3}{2}n+n\log n-\left \lfloor \frac{n}{2} \right \rfloor+3$   \\ \hline
\textbf{Ladner-Fisher}          &$3n-4+n\log n-4\left \lfloor \frac{n}{2} \right \rfloor$&     $\frac{3n}{4}-1+\frac{n\log n}{4}$ &                        2&$\log n+2$& $3n+\frac{n\log n}{2}-\left \lfloor \frac{n}{2} \right \rfloor+1$\\ \hline
\end{tabular}
\end{table}

After comparing all the proposed designs, we confirm that the Quantum Slansky + Strategy 2 is the optimal depth design choice in this work. This is attributed to its minimal Toffoli-Depth and fewer CNOT operations involved, especially considering that Slansky only requires P-CNOT operations during CNOT operations, without the need for G-CNOT operations.

\subsection{Comparative Analysis: Optimal Depth Adder Vs. Existing Quantum adders}
In this subsection, we assess our proposed Optimal Depth Adder against existing quantum adder designs, as detailed in Table \ref{tab:table_Compare_Overall}.

\begin{table*}[ht!]
\renewcommand\arraystretch{1.923} 
\caption{Performance analysis of different quantum adders.\\
The formula for $\omega(n)$ is $\omega(n)=n-\sum_{y=1}^\infty\left \lfloor\frac{n}{2^y}\right \rfloor$ and  $r$ represents the radix, with a range of $2 < r \leq n$.
\label{tab:table_Compare_Overall}}
\resizebox{1.0\textwidth}{!}{
\begin{tabular}{|cc|c|c|c|}
\hline
 \textbf{Adder}& \textbf{Year} & \textbf{Toffoli Count}
& \textbf{Toffoli Depth}& \textbf{Qubit Count}\\ \hline

 VBE RCA \cite{VBE}&$1995$&$4n-2$&$4n-2$&$3n+1$\\\hline

 Cuccaro RCA \cite{Cuccaro}&$2004$&$2n-1$&$2n-1$&$2n+2$\\\hline

 Draper In-place CLA \cite{Draper}&$2004$& \makecell{$10n-3\omega(n)-3\omega(n-1)$\\$-3\left \lfloor\log n\right \rfloor-3\left \lfloor\log (n-1)\right \rfloor-7$ }
 &\makecell{$8+ \left \lfloor\log n\right \rfloor+ \left \lfloor\log (n-1)\right \rfloor$\\$+ \left \lfloor\log \frac{n}3\right \rfloor+\left \lfloor\log \frac{n-1}3\right \rfloor$}&$4n-\omega(n)-\left \lfloor\log n\right \rfloor$ \\\hline

  Draper Out-of-place CLA \cite{Draper}& $2004$ &$5n-3\omega(n)-3\left \lfloor\log n\right \rfloor-1$ &$4+\left \lfloor\log n\right \rfloor+\left \lfloor\log \frac{n}3\right \rfloor$&$4n+1-\omega(n)-\left \lfloor\log n\right \rfloor$ \\\hline
  Takahashi Adder \cite{Takahashi08}&$2008$&$28n$&$30\log n$&$2n+\frac{3n}{\log n}$ \\\hline

 Takahashi RCA \cite{Takahashi09}&$2009$&$2n-1$&$2n-1$&$2n+1$\\\hline

 Takahashi Combination  \cite{Takahashi09}&$2009$&$7n$ &$18\log n$&$2n+\frac{3\cdot n}{\log n}$ \\\hline


 Wang RCA
 \cite{paper4_2016}&$2016$&$n$&$n$&$3n+1$\\ \hline

 Gidney RCA
 \cite{Gidney2018halvingcostof}&$2018$&$2n-2$&$n$&$3n-1$\\ \hline

 Gayathri RCA
 \cite{paper3_2021}&$2021$&$n$&$n$&$3n+1$\\ \hline


 Higher Radix Adder \cite{wang2023higher}&$2023$&\begin{tabular}[c]{@{}c@{}}$8n-\frac{n}{r}-(n-1)\pmod  r$\\$-3\omega(\frac{n}{r})-3\log n +3\log r -3$\end{tabular}
&\begin{tabular}[c]{@{}c@{}}$4 \log n+3r-2\log r$\\$ -2\log 3r+2\log (r-2)+2$\end{tabular}
 &\begin{tabular}[c]{@{}c@{}}$4n-\log n+\frac{n}{r}$\\$ -\omega(\frac{n}{r})+\log r-1$\end{tabular}\\\hline
 Quantum Ling Adder \cite{wang2023reducing}&$2023$&$13n -6\omega(\frac{n}{2})-6\lfloor \log{\frac{n}{2}} \rfloor-14$&$9+2\lfloor\log\frac{n}{2} \rfloor+2\lfloor\log\frac{n}{6} \rfloor$&$12n-6\omega(\frac{n}{2})-6\lfloor \log{\frac{n}{2}} \rfloor-10$\\\hline
 \rowcolor{cyan!6}  \textbf{Optimal Depth Adder+ Strategy 1}&$-$&$\frac{3}{2}n\log n +2\left \lceil \log n \right \rceil-n$&$2\log n+1$&$n+n\log n+\left \lceil \log n \right \rceil+2$\\\hline
 \rowcolor{cyan!6}  \textbf{Optimal Depth Adder+ Strategy 2}&$-$&$\frac{n\log n}{2}$&\textcolor{green_best}{$\log n+1$}&$n+n\log n+\left \lceil \log n \right \rceil+2$\\\hline
\end{tabular}}
\end{table*}

The proposed optimal Toffoli-Depth adder achieves the lowest Toffoli-Depth in the quantum computing, with a harmonious balance in Toffoli Count and Qubit Count. Specifically, our optimal depth adder with strategy 2 significantly reduces the computational time by halving the Toffoli-Depth to $\log n+1$, a roughly $50\%$ improvement over the best existing quantum adders. Following closely is our optimal adder with strategy 1, achieving a competitive Toffoli-Depth of $2\log n+1$.

\begin{figure}[ht!]
    \centering
    \subfigure[Toffoli-Depth Comparison\label{fig:TD}]{\includegraphics[width=0.92\linewidth]{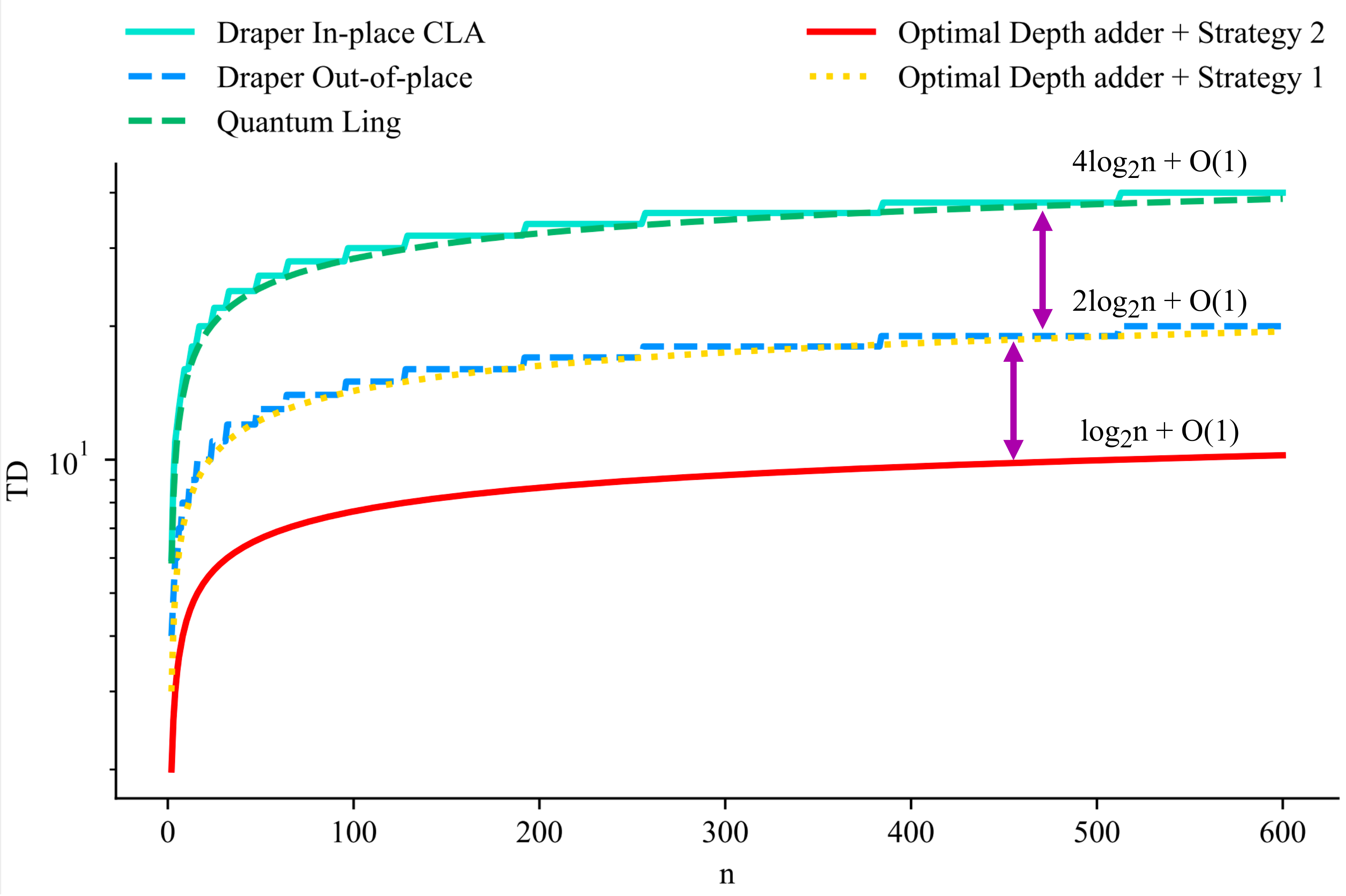}
    }
    \hfill
    \subfigure[Toffoli-Count Comparison\label{fig:TC}]{
    \includegraphics[width=.46\linewidth]{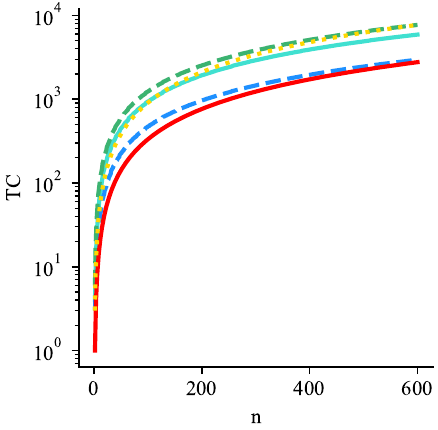}
    }
        \hfill
    \subfigure[Qubit-Count Comparison\label{fig:QC}]{
    \includegraphics[width=.46\linewidth]{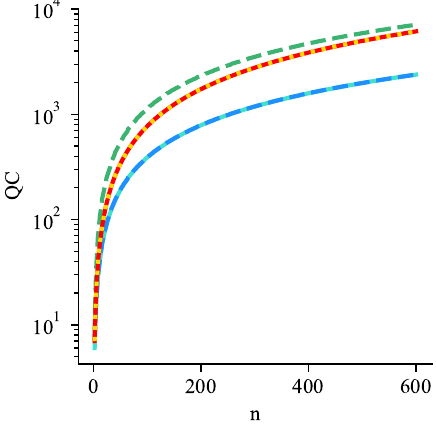}
    }
    \caption{Comparative Cost Analysis of Quantum Optimal Toffoli-Depth Adder and Other top 3 Prominent Quantum CLA Adders.  \label{fig:Comparison}}
\end{figure}

Interestingly, Figure \ref{fig:Comparison} illustrates that the quantum adders with the lowest Toffoli-Depth prior to our work, specifically the Draper In-place and Out-of-place CLAs\cite{Draper} and the Quantum Ling adder\cite{wang2023reducing}, demonstrate a higher Toffoli-Depth compared to our designs. Nevertheless, despite these existing work have slightly lower Toffoli-Count and Qubit-Count compared to our adders, they do not compensate for the longer time cost associated with their higher Toffoli-Depth.

Overall, our designs substantially enhance Toffoli-Depth while wisely managing quantum resources, thereby representing a considerable enhancement in both the efficiency and practicality over current quantum adders.

\subsection{Extension 1: Ling-based Optimal Toffoli-Depth adder}

\begin{figure}[ht!]
\centering
\includegraphics[width=0.82\linewidth]{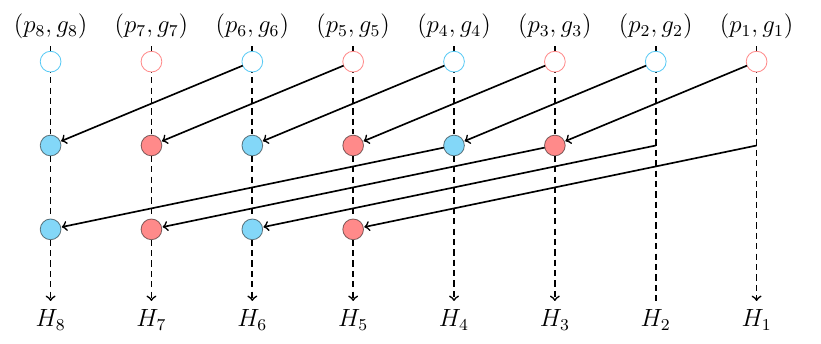}  
\caption{Kogge-Stone Prefix Tree using Ling Expansion.\label{fig:Classical_Ling_KS_Tree}}
\end{figure}

Inspired by the innovative quantum Ling adder proposed by Wang et al. \cite{wang2023reducing}, we introduce a novel quantum Ling structure based on the prefix tree adders proposed in this paper. In Figure~\ref{fig:Classical_Ling_KS_Tree}, a Ling-based adder using the quantum Kogge-Stone is presented as an example, which introduces two parallel prefix computation trees based on the quantum Kogge-Stone computation tree.

This design significantly reduces the Toffoli-Depth of the quantum Ling adder by half. However, this improvement is accompanied by an acceptable increase in Toffoli-Count and Qubit-Count. Specifically, the Toffoli-Depth of our Ling-based Kogge-Stone adder is $2\log {\frac{n}{2}}+8$, reflecting innovative adaptations in generation and propagation computations, thereby significantly optimizing the depth of the initial quantum Ling adder \cite{wang2023reducing}. This proposed extension structure demonstrates commendable performance in enhancing the efficiency of Ling-based quantum adders, thereby confirming the flexibility and portability of the adders proposed in our paper.

\begin{table*}[ht!]
\renewcommand\arraystretch{1.0} 
\caption{Performance analysis of Ling-based quantum Kogge-Stone adders utilizing Strategy 1.
\label{Ling_Strategy1}}

\begin{tabular}{|c|c|c|c|}
\hline
 \textbf{Adder} & \textbf{Toffoli Count}
& \textbf{Toffoli Depth}& \textbf{Qubit Count}\\ \hline
\textbf{ Quantum Ling Adder \cite{wang2023reducing}}&$13n -6\omega(\frac{n}{2})-6\lfloor \log{\frac{n}{2}} \rfloor-14$&$9+2\lfloor\log\frac{n}{2} \rfloor+2\lfloor\log\frac{n}{6} \rfloor$&$12n-6\omega(\frac{n}{2})-6\lfloor \log{\frac{n}{2}} \rfloor-10$\\\hline
 
 \textbf{K-S}&$3n\log n+n\log {\frac{n}{2}}-3n+5$&\textcolor{green_best}{$2\log n+2$}&$3n\log n -\frac{n}{2}+6$\\\hline

\textbf{K-S$+$ Ling}&$3n\log n+n$&$4\log n+6$&$3n\log n+2n\log {\frac{n}{2}}+\frac{n}{2}+3$\\\hline

\end{tabular}
\end{table*}

\begin{table}[ht!]
\centering
\caption{Performance analysis of Ling-based quantum Kogge-Stone adders utilizing Strategy 2.\label{Ling_Strategy2}}
\begin{tabular}{|c|c|c|c|c|c|}
\hline
\textbf{Adder} & \textbf{Extra T Count} & \textbf{Toffoli Count} & \textbf{Extra T Depth} & \textbf{Toffoli Depth} & \textbf{Qubit Count} \\ \hline
\textbf{K-S}&$8n\log n-14n+20$& $n\log {n}-1$ &          2&\textcolor{green_best}{$\log n+2$}&$3n\log n -\frac{n}{2}+6$\\ \hline
\textbf{K-S+Ling}  &$4n\log n-8n+8  $& $2n\log n -4n+4$&2 &$2\log n+8$&$3n\log n+2n\log {\frac{n}{2}}+\frac{n}{2}+3$\\ \hline
\end{tabular}
\end{table}

However, as shown in Table \ref{Ling_Strategy1} and Table \ref{Ling_Strategy2}, we also find that replacing traditional propagation and generation with Ling base is not beneficial for the prefix tree adders we propose. This is because the Ling structure introduces OR logic, requiring two Toffoli gates at each calculation node for propagation and generation.
Overall, using Ling base almost doubles the Toffoli-Depth of the prefix tree adders.

\subsection{Extension 2: Optimal Toffoli-depth Modular Adder\label{sec:Modular}}

\begin{figure*}[ht!]
    \centering
\includegraphics[scale=0.99]{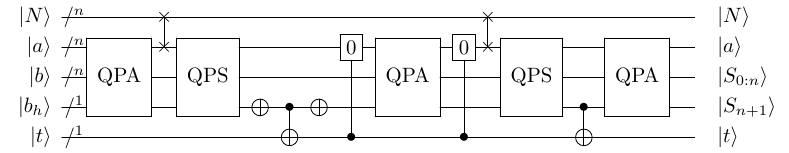}
\caption{Quantum VBE Modular Addition Framework~\cite{VBE}. Besides, the 'Set 0' gate in the framework can be implemented using several CNOTs. \label{Overall_Modular_Design}}
\end{figure*}

In this extension, we utilized the VBE modular addition framework mentioned in the paper ~\cite{VBE}, which incorporates the quantum prefix tree adders proposed in this paper, as illustrated in Figure \ref{Overall_Modular_Design}.

\begin{figure}[ht!]
    \centering
    \subfigure[Quantum Prefix Tree Adder \label{fig:Modular_Ling_adder}]{\includegraphics[width=0.46\linewidth]{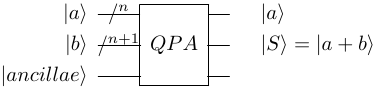}
    }
    \hspace{3mm}
    \subfigure[Quantum Prefix Tree Subtractor\label{fig:Modular_Ling_subtractor}]{\includegraphics[width=0.46\linewidth]{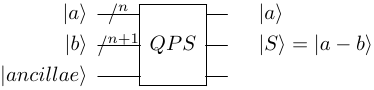}
    }
    \caption{Two important sub-modules of the proposed modular addition framework.}
\end{figure}

Our quantum modular addition framework primarily contains two important parts: the quantum prefix tree adder and the quantum prefix tree subtractor, with the latter being an altered version of the former. As illustrated in Figures \ref{fig:Modular_Ling_adder} and \ref{fig:Modular_Ling_subtractor}, these two structures are very similar to each other.

The implementation of the quantum subtractor is based on certain modifications to the quantum adder, as indicated by Equation \eqref{formula: sub}. Specifically, we begin to take the bit-wise complement of $a$, then add $b$ to obtain the result $a' + b$. Finally, we take the two's complement of the qubits of $a$ and the output qubits. Remarkably, this process can be easily accomplished using a series of quantum NOT gates. Hence, the cost of the quantum prefix tree subtractor is very close to the cost of the corresponding adder.
\begin{align}
&A-B=(A'+ B)'\label{formula: sub}
\end{align}

In Table \ref{tab:table_VBE}, we demonstrate the performance of our designed modular adders compared to other existing quantum modular adders. Our designs exhibits superior Toffoli-Depth, particularly the Quantum Sklansky with Strategy 2 design achieving an optimal Toffoli-Depth of $5\log n+5$. However, this achievement comes at the cost of a higher Qubit-Count and Toffoli-Count.

\begin{table*}[ht!]
\renewcommand\arraystretch{1.9} 
\caption{Performance analysis of different VBE-based quantum modular adders.\\
The formula for $\omega(n)$ is $\omega(n)=n-\sum_{y=1}^\infty\left \lfloor\frac{n}{2^y}\right \rfloor$, where $n$ denotes the bit-width of the addend.
\label{tab:table_VBE}}
\resizebox{1.0\textwidth}{!}{
\begin{tabular}{|c|c|c|c|c|}
\hline
 \textbf{Modular Adder}&\textbf{Year} &\textbf{Toffoli Count}
&\textbf{Toffoli Depth}&\textbf{Qubit Count}\\ \hline
 VBE\cite{VBE}
 &$1995$&$20n-10$&$20n-10$&$4n+2$\\\hline
 Cucarro \cite{Cuccaro}
 &$2004$&$10n-5$&$10n-5$&$3n+3$\\\hline
Draper In-place \cite{Draper}&$2004$& \makecell{$50n-15\omega(n)-15\omega(n-1)$\\$-15\left \lfloor\log n\right \rfloor-15\left \lfloor\log (n-1)\right \rfloor-35$ }
 &\makecell{$40+5\left \lfloor\log n\right \rfloor+5\left \lfloor\log (n-1)\right \rfloor$\\$+5\left \lfloor\log \frac{n}3\right \rfloor+5\left \lfloor\log \frac{n-1}3\right \rfloor$}&$5n-\omega(n)-\left \lfloor\log n\right \rfloor+1$ \\\hline

 \rowcolor{green!6} Brent-Kung+Strategy 1&-&$25n-15\omega(n)-15\left \lfloor\log n\right \rfloor-5$ &$20+5\left \lfloor\log n\right \rfloor+5\left \lfloor\log \frac{n}3\right \rfloor$&$5n+2-\omega(n)-\left \lfloor\log n\right \rfloor$ \\ \hline
 \rowcolor{green!6}Sklansky+Strategy 1 &-              &$\frac{15}{2}n\log n +10\left \lceil \log n \right \rceil-5n$&$10\log n+5$&$2n+n\log n+\left \lceil \log n \right \rceil+3$\\ \hline
 \rowcolor{green!6}Kogge-Stone+Strategy 1&-&$15n\log n+5n\log {\frac{n}{2}}-15n+25$&$10\log n+10$&$3n\log n +\frac{n}{2}+7$\\ \hline
 \rowcolor{green!6}Han-Carlson+Strategy 1  &-          &$5n+\frac{15}{2}n\log n -10\left \lfloor \frac{n}{2} \right \rfloor$&$10\log n+15$&$\frac{5}{2}n+n\log n-\left \lfloor \frac{n}{2} \right \rfloor+4$   \\ \hline
 \rowcolor{green!6}  Ladner-Fisher+Strategy 1 &-         & $\frac{65n}{4}+\frac{15n\log n}{4}-10\left \lfloor \frac{n}{2} \right \rfloor-15$&$10\log n+15$& $4n+\frac{n\log n}{2}-\left \lfloor \frac{n}{2} \right \rfloor+2$\\ \hline

 \rowcolor{blue!6}   Brent-Kung+Strategy 2&-& $10n-5\log n -10 $ &$10\log n-5$&$5n+2-\omega(n)-\left \lfloor\log n\right \rfloor$ \\ \hline
 \rowcolor{blue!6}  Sklansky+Strategy 2 &-              &$\frac{5n\log n}{2}$&\textcolor{green_best}{$5\log n+5$}&$2n+n\log n+\left \lceil \log n \right \rceil+3$\\ \hline
 \rowcolor{blue!6}  Kogge-Stone+Strategy 2&-&$5n\log {n}-5$&$5\log n+10$&$3n\log n +\frac{n}{2}+7$\\ \hline
 \rowcolor{blue!6}  Han-Carlson +Strategy 2 &-          &$\frac{5n\log n}{2}$&$5\log n+10$&$\frac{5}{2}n+n\log n-\left \lfloor \frac{n}{2} \right \rfloor+4$   \\ \hline
 \rowcolor{blue!6}  Ladner-Fisher+Strategy 2 &-         &$\frac{15n}{4}-5+\frac{5n\log n}{4}$&$5\log n+10$& $4n+\frac{n\log n}{2}-\left \lfloor \frac{n}{2} \right \rfloor+2$\\ \hline
\end{tabular}}
\end{table*}

Above all, our design achieves optimal depth in the quantum world, maintaining an exceptionally low Toffoli-Depth. This accomplishment provides a crucial foundation for efficient arithmetic in quantum computing, unlocking new possibilities for high-performance quantum algorithms and enhancing the overall scalability and practicality of quantum computing. 
\section{Conclusion}\label{sec: Conc}

In conclusion, this paper has presented a novel architecture for enhancing the efficiency of quantum adders. By integrating the different prefix tree  into the quantum domain, we have achieved a groundbreaking optimal Toffoli-Depth quantum adder, which marks a parallel to the development of classical adder analysis. 
Furthermore, our work includes a significant quantum Ling expansion, which substantially improves the quantum Ling adder's performance by using our innovative structure. Moreover, our work also incorporates a modular addition extension.
The proposed quantum optimal depth adder achieves a Toffoli-Depth of $\log{}n + \mathcal{O}(1)$, which represents a significant improvement over all the previous quantum adders, which had a Toffoli-Depth of at least $2 \log{}n + \mathcal{O}(1)$. 
Our work marks a significant advancement in the field of quantum computing, addressing the challenge of depth optimization in the area of quantum addition.

In future, several research directions hold substantial promise. For instance, one of these directions is exploring the optimal adder structure on the real quantum computers. Moreover, Quantum Error Correction (QEC) also plays a crucial role in improving the adder's fault tolerance and reliability. 

\section*{Code Availability}
The relevant code will be available as a public repository online upon this paper's acceptance.

\begin{acks}
This research is supported by the National Research Foundation, Singapore under its Quantum Engineering Programme Initiative. Any opinions, findings and conclusions or recommendations expressed in this material are those of the authors and do not reflect the views of National Research Foundation, Singapore.
\end{acks}

\bibliographystyle{ACM-Reference-Format}
\bibliography{main}


\begin{thebibliography}{18}


\ifx \showCODEN    \undefined \def \showCODEN     #1{\unskip}     \fi
\ifx \showDOI      \undefined \def \showDOI       #1{#1}\fi
\ifx \showISBNx    \undefined \def \showISBNx     #1{\unskip}     \fi
\ifx \showISBNxiii \undefined \def \showISBNxiii  #1{\unskip}     \fi
\ifx \showISSN     \undefined \def \showISSN      #1{\unskip}     \fi
\ifx \showLCCN     \undefined \def \showLCCN      #1{\unskip}     \fi
\ifx \shownote     \undefined \def \shownote      #1{#1}          \fi
\ifx \showarticletitle \undefined \def \showarticletitle #1{#1}   \fi
\ifx \showURL      \undefined \def \showURL       {\relax}        \fi
\providecommand\bibfield[2]{#2}
\providecommand\bibinfo[2]{#2}
\providecommand\natexlab[1]{#1}
\providecommand\showeprint[2][]{arXiv:#2}

\bibitem[Brent and Kung(2004)]%
        {BK}
\bibfield{author}{\bibinfo{person}{Richard Brent} {and} \bibinfo{person}{Hsiang Kung}.} \bibinfo{year}{2004}\natexlab{}.
\newblock \showarticletitle{A Regular Layout for Parallel Adders}.
\newblock \bibinfo{journal}{\emph{IEEE Trans. Comput.}}  \bibinfo{volume}{31} (\bibinfo{date}{06} \bibinfo{year}{2004}).
\newblock
\urldef\tempurl%
\url{https://doi.org/10.1109/TC.1982.1675982}
\showDOI{\tempurl}


\bibitem[Cuccaro et~al\mbox{.}(2004)]%
        {Cuccaro}
\bibfield{author}{\bibinfo{person}{Steven~A. Cuccaro}, \bibinfo{person}{Thomas~G. Draper}, \bibinfo{person}{Samuel~A. Kutin}, {and} \bibinfo{person}{David~Petrie Moulton}.} \bibinfo{year}{2004}\natexlab{}.
\newblock \bibinfo{title}{A new quantum ripple-carry addition circuit}.
\newblock
\newblock
\showeprint[arxiv]{quant-ph/0410184}~[quant-ph]


\bibitem[Draper et~al\mbox{.}(2004)]%
        {Draper}
\bibfield{author}{\bibinfo{person}{Thomas Draper}, \bibinfo{person}{Samuel Kutin}, \bibinfo{person}{Eric Rains}, {and} \bibinfo{person}{Krysta Svore}.} \bibinfo{year}{2004}\natexlab{}.
\newblock \showarticletitle{A logarithmic-depth quantum carry-lookahead adder}.
\newblock \bibinfo{journal}{\emph{Quantum Information and Computation}}  \bibinfo{volume}{6} (\bibinfo{date}{07} \bibinfo{year}{2004}).
\newblock
\urldef\tempurl%
\url{https://doi.org/10.26421/QIC6.4-5-4}
\showDOI{\tempurl}


\bibitem[Gidney(2018)]%
        {Gidney2018halvingcostof}
\bibfield{author}{\bibinfo{person}{Craig Gidney}.} \bibinfo{year}{2018}\natexlab{}.
\newblock \showarticletitle{Halving the cost of quantum addition}.
\newblock \bibinfo{journal}{\emph{{Quantum}}}  \bibinfo{volume}{2} (\bibinfo{date}{June} \bibinfo{year}{2018}), \bibinfo{pages}{74}.
\newblock
\showISSN{2521-327X}
\urldef\tempurl%
\url{https://doi.org/10.22331/q-2018-06-18-74}
\showDOI{\tempurl}


\bibitem[Held and Spirkl(2017)]%
        {Best_Classical_Adder}
\bibfield{author}{\bibinfo{person}{Stephan Held} {and} \bibinfo{person}{Sophie~Theresa Spirkl}.} \bibinfo{year}{2017}\natexlab{}.
\newblock \showarticletitle{Binary Adder Circuits of Asymptotically Minimum Depth, Linear Size, and Fan-Out Two}.
\newblock \bibinfo{journal}{\emph{ACM Trans. Algorithms}} \bibinfo{volume}{14}, \bibinfo{number}{1}, Article \bibinfo{articleno}{4} (\bibinfo{date}{dec} \bibinfo{year}{2017}), \bibinfo{numpages}{18}~pages.
\newblock
\showISSN{1549-6325}
\urldef\tempurl%
\url{https://doi.org/10.1145/3147215}
\showDOI{\tempurl}


\bibitem[Kogge and Stone(1973)]%
        {KSTree}
\bibfield{author}{\bibinfo{person}{Peter~M. Kogge} {and} \bibinfo{person}{Harold~S. Stone}.} \bibinfo{year}{1973}\natexlab{}.
\newblock \showarticletitle{A Parallel Algorithm for the Efficient Solution of a General Class of Recurrence Equations}.
\newblock \bibinfo{journal}{\emph{IEEE Trans. Comput.}} \bibinfo{volume}{C-22}, \bibinfo{number}{8} (\bibinfo{year}{1973}), \bibinfo{pages}{786--793}.
\newblock
\urldef\tempurl%
\url{https://doi.org/10.1109/TC.1973.5009159}
\showDOI{\tempurl}


\bibitem[Lim et~al\mbox{.}(2023)]%
        {Payoff_Function}
\bibfield{author}{\bibinfo{person}{Sejin Lim}, \bibinfo{person}{Hyunjun Kim}, \bibinfo{person}{Kyungbae Jang}, \bibinfo{person}{Siyi Wang}, \bibinfo{person}{Anubhab Baksi}, \bibinfo{person}{Anupam Chattopadhyay}, {and} \bibinfo{person}{Hwajeong Seo}.} \bibinfo{year}{2023}\natexlab{}.
\newblock \showarticletitle{Optimized Quantum Circuit Implementation of Payoff Function}. \bibinfo{pages}{1--6}.
\newblock
\urldef\tempurl%
\url{https://doi.org/10.1109/VLSI-SoC57769.2023.10321843}
\showDOI{\tempurl}


\bibitem[Ling(1966)]%
        {Ling}
\bibfield{author}{\bibinfo{person}{Huei Ling}.} \bibinfo{year}{1966}\natexlab{}.
\newblock \showarticletitle{High Speed Binary Parallel Adder}.
\newblock \bibinfo{journal}{\emph{IEEE Transactions on Electronic Computers}} \bibinfo{volume}{EC-15}, \bibinfo{number}{5} (\bibinfo{year}{1966}), \bibinfo{pages}{799--802}.
\newblock
\urldef\tempurl%
\url{https://doi.org/10.1109/PGEC.1966.264571}
\showDOI{\tempurl}


\bibitem[S~S et~al\mbox{.}(2021)]%
        {paper3_2021}
\bibfield{author}{\bibinfo{person}{Gayathri S~S}, \bibinfo{person}{R. Kumar}, \bibinfo{person}{Dhanalakshmi Samiappan}, \bibinfo{person}{Brajesh~Kumar Kaushik}, {and} \bibinfo{person}{Majid Haghparast}.} \bibinfo{year}{2021}\natexlab{}.
\newblock \showarticletitle{T-Count Optimized Wallace Tree Integer Multiplier for Quantum Computing}.
\newblock \bibinfo{journal}{\emph{International Journal of Theoretical Physics}}  \bibinfo{volume}{60} (\bibinfo{date}{08} \bibinfo{year}{2021}), \bibinfo{pages}{1--13}.
\newblock
\urldef\tempurl%
\url{https://doi.org/10.1007/s10773-021-04864-3}
\showDOI{\tempurl}


\bibitem[Shor(1996)]%
        {Shor}
\bibfield{author}{\bibinfo{person}{Peter Shor}.} \bibinfo{year}{1996}\natexlab{}.
\newblock \showarticletitle{Algorithms for Quantum Computation: Discrete Logarithms and Factoring}.
\newblock \bibinfo{journal}{\emph{Proceedings of 35th Annual Symposium on Foundations of Computer Science}} (\bibinfo{date}{10} \bibinfo{year}{1996}).
\newblock
\urldef\tempurl%
\url{https://doi.org/10.1109/SFCS.1994.365700}
\showDOI{\tempurl}


\bibitem[Sklansky(1960)]%
        {Sklansky}
\bibfield{author}{\bibinfo{person}{J. Sklansky}.} \bibinfo{year}{1960}\natexlab{}.
\newblock \showarticletitle{Conditional-Sum Addition Logic}.
\newblock \bibinfo{journal}{\emph{Electronic Computers, IRE Transactions on}}  \bibinfo{volume}{EC-9} (\bibinfo{date}{07} \bibinfo{year}{1960}), \bibinfo{pages}{226 -- 231}.
\newblock
\urldef\tempurl%
\url{https://doi.org/10.1109/TEC.1960.5219822}
\showDOI{\tempurl}


\bibitem[Takahashi and Kunihiro(2008)]%
        {Takahashi08}
\bibfield{author}{\bibinfo{person}{Yasuhiro Takahashi} {and} \bibinfo{person}{Noboru Kunihiro}.} \bibinfo{year}{2008}\natexlab{}.
\newblock \showarticletitle{A fast quantum circuit for addition with few qubits}.
\newblock \bibinfo{journal}{\emph{Quantum Information and Computation}}  \bibinfo{volume}{8} (\bibinfo{date}{07} \bibinfo{year}{2008}), \bibinfo{pages}{636--649}.
\newblock
\urldef\tempurl%
\url{https://doi.org/10.26421/QIC8.6-7-5}
\showDOI{\tempurl}


\bibitem[Takahashi et~al\mbox{.}(2009)]%
        {Takahashi09}
\bibfield{author}{\bibinfo{person}{Yasuhiro Takahashi}, \bibinfo{person}{Seiichiro Tani}, {and} \bibinfo{person}{Noboru Kunihiro}.} \bibinfo{year}{2009}\natexlab{}.
\newblock \showarticletitle{Quantum Addition Circuits and Unbounded Fan-Out}.
\newblock \bibinfo{journal}{\emph{Quantum Information and Computation}}  \bibinfo{volume}{10} (\bibinfo{date}{10} \bibinfo{year}{2009}).
\newblock
\urldef\tempurl%
\url{https://doi.org/10.26421/QIC10.9-10-12}
\showDOI{\tempurl}


\bibitem[Vedral et~al\mbox{.}(1995)]%
        {VBE}
\bibfield{author}{\bibinfo{person}{Vlatko Vedral}, \bibinfo{person}{Adriano Barenco}, {and} \bibinfo{person}{Artur Ekert}.} \bibinfo{year}{1995}\natexlab{}.
\newblock \showarticletitle{Quantum Networks for Elementary Arithmetic Operations}.
\newblock \bibinfo{journal}{\emph{Physical Review A}}  \bibinfo{volume}{54} (\bibinfo{date}{11} \bibinfo{year}{1995}).
\newblock
\urldef\tempurl%
\url{https://doi.org/10.1103/PhysRevA.54.147}
\showDOI{\tempurl}


\bibitem[Wang et~al\mbox{.}(2016)]%
        {paper4_2016}
\bibfield{author}{\bibinfo{person}{Feng Wang}, \bibinfo{person}{Mingxing Luo}, \bibinfo{person}{Huiran Li}, \bibinfo{person}{Zhiguo Qu}, {and} \bibinfo{person}{Xiaojun Wang}.} \bibinfo{year}{2016}\natexlab{}.
\newblock \showarticletitle{Improved quantum ripple-carry addition circuit}.
\newblock \bibinfo{journal}{\emph{Science China Information Sciences}}  \bibinfo{volume}{59} (\bibinfo{date}{02} \bibinfo{year}{2016}).
\newblock
\urldef\tempurl%
\url{https://doi.org/10.1007/s11432-015-5411-x}
\showDOI{\tempurl}


\bibitem[Wang et~al\mbox{.}(2023)]%
        {wang2023higher}
\bibfield{author}{\bibinfo{person}{Siyi Wang}, \bibinfo{person}{Anubhab Baksi}, {and} \bibinfo{person}{Anupam Chattopadhyay}.} \bibinfo{year}{2023}\natexlab{}.
\newblock \showarticletitle{A Higher Radix Architecture for Quantum Carry-lookahead Adder}.
\newblock \bibinfo{journal}{\emph{Scientific Reports}}  \bibinfo{volume}{13} (\bibinfo{date}{09} \bibinfo{year}{2023}).
\newblock
\urldef\tempurl%
\url{https://doi.org/10.1038/s41598-023-41122-4}
\showDOI{\tempurl}


\bibitem[Wang and Chattopadhyay(2023)]%
        {wang2023reducing}
\bibfield{author}{\bibinfo{person}{Siyi Wang} {and} \bibinfo{person}{Anupam Chattopadhyay}.} \bibinfo{year}{2023}\natexlab{}.
\newblock \showarticletitle{Reducing Depth of Quantum Adder using Ling Structure}. In \bibinfo{booktitle}{\emph{VLSI-SoC}}.
\newblock


\bibitem[Wang et~al\mbox{.}(2024)]%
        {divider}
\bibfield{author}{\bibinfo{person}{Siyi Wang}, \bibinfo{person}{Eugene Lim}, {and} \bibinfo{person}{Anupam Chattopadhyay}.} \bibinfo{year}{2024}\natexlab{}.
\newblock \bibinfo{title}{Boosting the Efficiency of Quantum Divider through Effective Design Space Exploration}.
\newblock
\newblock
\showeprint[arxiv]{2403.01206}~[quant-ph]


\end{thebibliography}
\end{document}